\documentclass[aps,prl,twocolumn]{revtex4}
\usepackage{amsmath}
\usepackage{graphics}
\usepackage{graphicx}
\usepackage{color}

\begin{document}

\title{Large Broadening of the Superconducting Transition by Fluctuations in a 3D
Metal at High Magnetic Fields: The MgB$_{2}$ case. }
\author{T. Maniv and V. Zhuravlev}
\date{\today}

\begin{abstract}
It is shown that the transition to the low temperature superconducting state
in a 3D metal at high magnetic field is smeared dramatically by thermal
fluctuation of the superconducting order parameter. \ The resulting
superconducting-to-normal crossover occurs in a vortex liquid state which is
extended well below the mean-field $H_{c2}$. Application to MgB$_{2}$ yields
good quantitative agreement with recently reported data of dHvA oscillation
in the superconducting state.
\end{abstract}

\affiliation{Chemistry Department, Technion-Israel Institute of
Technology, Haifa 32000, Israel }

\maketitle

It is well known that the transition from the normal to the superconducting
(SC) state in type-II 3D superconductors in the absence of external magnetic
field is a sharp, second-order phase transition, with a vanishing order
parameter at the transition temperature $T_{c}$ continuously growing with
the decreasing temperature below $T_{c}$. \ Fluctuations effect can smear
the transition significantly in high $T_{c}$ and low-dimensional
superconductors \cite{Kes91}, where the phase space accessible for the
fluctuations is dramatically enhanced. The influence of an external magnetic
field is similar to an effective reduction of dimensionality\cite{Lee72},
resulting in a significant smearing of the transition even at very low
temperatures. Such strong smearing effects have been observed in varies
quasi 2D low $T_{c}$ superconductors at high magnetic fields \cite{Urbach92}%
, \cite{Sasaki03},\cite{Ito99}. \ 

In the present paper we show theoretically, and confirm by comparison with
very recent de-Haas van-Alphen (dHvA) oscillation data in the SC state \cite
{Fletcher04}, that the smearing of the SC transition by fluctuations in a
conventional 3D type-II superconductor, such as MgB$_{2}$, at high magnetic
fields, is surprisingly strong, comparable in magnitude to that in 2D
superconductors. This conclusion is reached by generalizing the Bragg-chain
model of the 2D vortex liquid state at high perpendicular magnetic field\cite
{RMP01} to an array of strongly coupled 2D SC layers. Within this model it
is found that, similar to the situation in a single 2D SC layer, the vortex
lattice melting point in a 3D superconductor at low temperature $T$ is
located well below the mean-field (MF) upper critical field $H_{c2}\left(
T\right) $, so that in a broad field range above the melting point the
corresponding system of fluctuations is equivalent to a 1D array of SC
quantum dots at zero magnetic field \cite{Ebner}.\ 

Our starting point is the microscopic BCS Hamiltonian for electrons in a
layered 3D metal, interacting via an effective two-body attractive
potential, under the influence of a strong static magnetic field. We assume,
for simplicity, that the magnetic field direction (along $z$-axis) is
perpendicular to the layers situated in $(x,y)$-plane. Writing down the
functional integral expression for the partition function of this system,
the electronic field can be eliminated by introducing bosonic complex field $%
\Delta \left( \mathbf{r}\right) $ ( by means of the Hubbard-Stratonovich
transformations), which describes all possible realizations of Cooper-pairs
condensates \cite{Tesanovic94},\cite{RMP01}. Expansion of the resulting free
energy functional, $F_{G}\left[ \Delta \left( \mathbf{r}\right) \right] $,
in the order parameter up to the quartic term is a good approximation for
magnetic fields around mean field $H_{c2}$.

In the lowest Landau level approximation, which is valid at high magnetic
fields and low temperatures \cite{Tesanovic92}, the most general form of the
order parameter $\Delta \left( \mathbf{r}\right) $ is a coherent
superposition of Landau wave functions, $\phi _{q}\left( x,y\right) =\exp %
\left[ iqx-\left( y/a_{H}+qa_{H}/2\right) ^{2}\right] $: \ 
\begin{equation}
\Delta \left( x,y,z\right) =\sum_{n,m}c_{q}\left( z\right) \phi _{q}\left(
x,y\right)  \label{odpar}
\end{equation}
where $c_{q}\left( z\right) $ are arbitrary complex functions of the
coordinate $z$ , defined on the quasi continuous lattice $q=\frac{2\pi }{%
a_{x}a_{H}}\left( n+m/\sqrt{N}\right) $, $n,m=-\sqrt{N}/2+1,\dots ,\sqrt{N}%
/2 $, which determines the projections of the orbital guiding centers on the 
$y$ axis. Here $N$ is the total number of flux lines threading the SC
sample, and $a_{H}=\sqrt{c\hbar /eH}$ is the magnetic length. \ In the case
when all the coefficients $c_{q}$ are different from zero there is one to
one correspondence between all $N$ guiding centers and their projections on
the $y$ axis.

As discussed in detail in Ref.\cite{ZMPRB02}, such a quasi continuous
configuration costs a large fraction of the total SC condensation energy,
and may be therefore omitted at magnetic fields well below MF $H_{c2}$. \ In
this region it is sufficient to use in Eq.(\ref{odpar}) a discrete subset of 
$\sqrt{N}$ Landau orbitals $\phi _{q_{n}}\left( x,y\right) $, with $q_{n}=%
\frac{2\pi }{a_{x}a_{H}}n$, $n=0,\pm 1,\pm 2,...$ . Including the omitted
high energy modes would always increase the fluctuation effect. Thus, near
and above the MF transition the fluctuation effect, calculated in the
discrete chain approximation, should always underestimate the observed
effect. On the other hand, an upper bound on the fluctuation effect can be
obtained from the 2D model \cite{ManivPhysB02}, where the discrete chain
representation yields reasonably good agreement with exact numerical
simulation \cite{Kato93}.

Writing $c_{q_{n}}\left( z\right) \equiv c_{n}\left( z\right) =$ $\left|
c_{n}\left( z\right) \right| e^{i\varphi _{n}\left( z\right) }$ the phase $%
\varphi _{n}\left( z\right) $ determines the relative lateral position $%
x_{n}=-\varphi _{n}/q_{n}$ of the $n$-th Landau orbital within a single 2D
layer. \ It can be readily shown that, by selecting the $x$ axis along the
principal crystallographic axis of the 2D Abrikosov vortex lattice, the
variables $\xi _{n}\equiv \varphi _{n}-\varphi _{n-1}$ describe the lateral
positions of the most easily sliding vortex chains in the vortex lattice,
which are generated mainly by interference between two neighboring Landau
orbitals \cite{ZMPRB02}.\ 

The corresponding free energy functional, projected on the subspace of the
lowest Landau level, can be written in a local anisotropic Ginzburg-Landau
(GL) form \cite{RMP01},\cite{ZMunpub}: 
\begin{equation}
F_{GL}=\int \frac{d^{3}r}{\upsilon}\left[ -\alpha \left| \Delta \left( 
\mathbf{r}\right) \right| ^{2}+\frac{\beta}{2} \left| \Delta \left( \mathbf{r%
}\right) \right| ^{4}+\gamma \left| \frac{\partial \Delta \left( \mathbf{r}%
\right) }{\partial z}\right| ^{2}\right]  \label{pf1}
\end{equation}
where $\upsilon =\pi a_{H}^{2}d$ is the volume of a single vortex per layer
and the effective GL coefficients $\alpha $,$\beta $, and $\gamma $,\ can be
expressed in terms of the microscopic normal electron parameters (see
below). \ It is convenient to divide the length $L_{z}$ of the sample
perpendicular to the layers into $N_{z}$ segments of length $d$ , $%
L_{z}=N_{z}d$ , where $d$ is the interlayer distance, and define a discrete
set $c_{n,\zeta }=\left| c_{n,\zeta }\right| e^{i\varphi _{n,\zeta }}\equiv
c_{n}\left( z_{\zeta }\right) $ , $\zeta =1,2,...N_{z}$ , with the periodic
boundary conditions: \ $c_{n,N_{z}+1}=c_{n,1}$.

Using Eq.(\ref{odpar}) the partition function, 
\begin{equation*}
\mathcal{Z}\equiv \mathcal{Z}_{ch}^{\sqrt{N}}=\int D\Delta \left( \mathbf{r}%
\right) D\Delta ^{\star }\left( \mathbf{r}\right) \exp \left\{ -F_{GL}\left[
\Delta \left( \mathbf{r}\right) \right] /k_{B}T\right\}
\end{equation*}
where $\mathcal{Z}_{ch}$ is evaluated as a multiple integral: $%
\prod_{n,\zeta }\int |c_{n,\zeta }|d|c_{n,\zeta }|\int d\varphi _{n,\zeta
}e^{-F_{GL}/(\sqrt{N}k_{B}T)}$, with 
\begin{eqnarray}
&&\frac{F_{GL}}{\left( a_{x}/\sqrt{2\pi }\right) \sqrt{N}}=-\alpha
\sum_{n,\zeta }\left| c_{n,\zeta }\right| ^{2}+\eta \sum_{n,\zeta }\left|
c_{n,\zeta +1}-c_{n,\zeta }\right| ^{2}  \notag \\
&&+ \frac{\beta}{2^{3/2}} \sum_{n,s,p;\zeta }\lambda
^{s^{2}+p^{2}}|c_{n,\zeta }||c_{n+s+p,\zeta }||c_{n+s,\zeta }||c_{n+p,\zeta
}|  \notag \\
&&\times e^{i\left( \varphi _{n+s,\zeta }+\varphi _{n+p,\zeta }-\varphi
_{n,\zeta }-\varphi _{n+s+p,\zeta }\right) }  \label{fren}
\end{eqnarray}
and $\eta =d^{2}\gamma $. In this Landau orbital representation of the GL
functional, Eq.(\ref{pf1}), the off-diagonal elements constitute a rapidly
convergent series in the small expansion parameter $\lambda \equiv
e^{-\left( \pi /a_{x}\right) ^{2}}\approx 0.066$, where all terms of order
higher than the second may be neglected.

For the 3D electronic band structure under study here the characteristic
interlayer Josephson tunneling amplitude $\eta $ is much larger than the
minimal intra-layer shear stiffness, $4\lambda ^{2}\sim 10^{-2}$, \
characterizing the principal axis, $x$, in the vortex lattice. \ Under this
condition the low energy fluctuations of $\Delta \left( \mathbf{r}\right) $
correspond to collectively sliding chains of vortices in different layers,
such that the corresponding fluctuating magnetic flux lines remain nearly
parallel to each other (and so to the external magnetic field). Similar to
the situation in a 2D superconductor \cite{RMP01}, these low-lying shear
fluctuations determine the vortex lattice melting point to be well below $%
H_{c2}$. \ 

For magnetic fields $H$ above this melting point the second order terms in $%
\lambda $, which oscillate as functions of the phases $\varphi _{n,\zeta }$,
are averaged to zero by the integrations over $\varphi _{n,\zeta }$, and the
effective free energy functional $F_{GL}$ can be written in the very simple,
independent vortex chain form $F_{GL}=\sqrt{N}\sum_{n}f_{GL}^{n}$ , with 
\begin{eqnarray}
f_{GL}^{n} &=&\frac{a_{x}}{\sqrt{2\pi }}\sum_{\zeta }\left\{ -\alpha \left|
c_{n,\zeta }\right| ^{2}+\frac{\beta }{2^{3/2}}\left| c_{n,\zeta }\right|
^{4}\right.  \notag \\
&+&\left. \eta \left| c_{n,\zeta +1}-c_{n,\zeta }\right| ^{2}\right\} ,
\label{fGL^n}
\end{eqnarray}
which may be considered as an effective GL energy functional for a single
vortex line. In this expression we have also neglected the first order terms
in $\lambda $, since they effectively yield a small additive correction to $%
\beta $, so that their influence on the critical behavior is unimportant 
\cite{ManivPhysB02}.

The calculation of the corresponding partition function is rather
straightforward. The phase variables, $\varphi _{n,\zeta }$, can be readily
integrated out. The resulting expression can be written as a functional
integral over the squared amplitude $y_{\zeta }\equiv \left( \beta
a_{x}/2k_{B}T\sqrt{\pi }\right) ^{1/2}\left| c_{\zeta }\right| ^{2}$ , with
an effective GL functional, incorporating phase fluctuations: 
\begin{eqnarray}
f_{GL,eff}&=&k_{B}T\sum_{\zeta }\Big( -\sqrt{2}xy_{\zeta }+\frac{1}{2}%
y_{\zeta }^{2}+2\kappa y_{\zeta }  \notag \\
&-& \ln I_{0}\left( 2\kappa \sqrt{y_{\zeta}y_{\zeta +1}}\right) \Big)
\label{xkapbeta} \\
x\equiv \frac{\alpha}{\sqrt{2\beta \beta _{a}k_{B}T}}&;&\quad \kappa \equiv 
\frac{ \eta }{\sqrt{\beta \beta _{a}k_{B}T}};\qquad \beta _{a}\equiv \frac{%
\sqrt{\pi }}{a_{x}}  \notag
\end{eqnarray}
Here and below, for the sake of notation simplicity, we drop the vortex
chain indices.

We can then estimate the partition function in two limiting situations, for
weak ($\kappa \ll 1$) and strong ($\kappa \gg 1$) interlayer coupling. In
the former limit the integrals over amplitudes can be calculated explicitly
with the well known result \cite{RMP01},\cite{ManivPhysB02}. It is
convenient to define the partition function per single vortex per layer, $%
\mathcal{Z}_{v}\equiv \mathcal{Z}^{1/\mathcal{N}}$ ($\mathcal{N}=NN_{z}$),
which is given by: 
\begin{equation}
\ln \frac{\mathcal{Z}_{v}}{\mathcal{Z}_{0}}=x^{2}+\ln \mathop{\rm erfc}%
\left( -x\right)  \label{lnzweak}
\end{equation}
where $\mathop{\rm erfc}(x)\equiv \frac{2}{\sqrt{\pi }}\int_{x}^{\infty
}e^{-y^{2}}dy$. \ For strong interlayer coupling the result is obtained by
using the steepest descend integration, leading to 
\begin{eqnarray}
&&\ln \frac{\mathcal{Z}_{v}}{\mathcal{Z}_{0}^{\prime }}=\sqrt{2}xy_{0}-\frac{%
1}{2}y_{0}^{2}-\frac{1}{2}\ln \left( 4\pi \kappa \right)  \notag \\
&&-\ln \frac{\left( x^{2}+1\right) ^{1/4}+\sqrt{\left( x^{2}+1\right)
^{1/2}+2^{1/2}\kappa }}{2^{3/4}}  \label{lnzstrong}
\end{eqnarray}
where $y_{0}=\left( x+\sqrt{x^{2}+1}\right) /\sqrt{2}$. In Eqs.(\ref{lnzweak}%
,\ref{lnzstrong}) $\mathcal{Z}_{0}$ and $\mathcal{Z}_{0}^{\prime }$ are
constants (i.e. independent of both $x$ and $\kappa $), and so are
thermodynamically unimportant.

\begin{figure}[tbp]
\includegraphics[width=9.5cm]{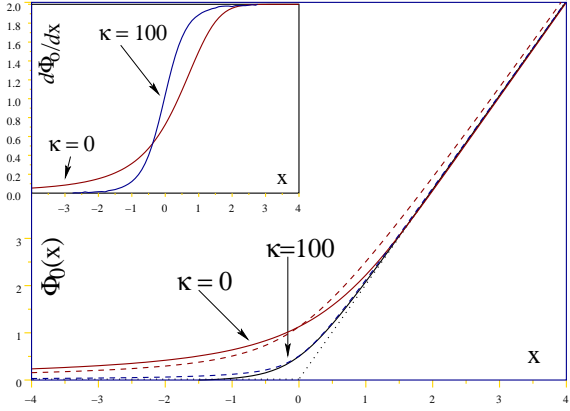} 
\caption{The spatially averaged mean-squared order parameter, measured in
units of $\protect\alpha _{0I}$ , as a function of the scaling parameter $x=%
\protect\alpha /\protect\sqrt{2\protect\beta \protect\beta _{a}k_{B}T}$ (see
text ) around the mean-field transition point $x=0$ in the weak and strong
coupling limits (Solid lines). Dashed lines represent the corresponding
results calculated from a simple interpolation formula. The doted straight
line represents the result of mean-field theory. Inset: The derivative with
respect to $x$ is plotted to emphasize the crossover region. }
\label{mgb2fig1}
\end{figure}

Thus, in the liquid state above the vortex lattice melting point one can
readily derive simple limiting expressions for the spatially averaged mean
square order parameter, $\left\langle \overline{\left| \Delta \right| ^{2}}%
\right\rangle \equiv \int d^{3}r\left\langle \left| \Delta \left( 
\overrightarrow{r}\right) \right| ^{2}\right\rangle /\left( \mathcal{N}%
\upsilon\right) $, in the general from: 
\begin{equation}
\left\langle \overline{\left| \Delta \right| ^{2}}\right\rangle =\frac{k_{B}T%
}{\mathcal{N}}\frac{\partial \ln \mathcal{Z}}{\partial \alpha }=\sqrt{\frac{%
k_{B}T}{2 \beta \beta _{a}}}\left( \frac{\partial \ln \mathcal{Z}_{v}}{%
\partial x}\right) \equiv \alpha _{0I}\Phi _{0}(x;\kappa )  \label{AvMSOP}
\end{equation}
where $\alpha _{0I}=\sqrt{\frac{k_{B}T}{2 \beta \beta _{a}}}$. The function $%
\Phi _{0}(x;\kappa )\equiv \frac{\partial \ln \mathcal{Z}_{v}}{\partial x}$
for $\kappa =0$ and $\kappa=100$ is shown in Fig. (\ref{mgb2fig1}). With
increasing interlayer coupling the mean-square order parameter at the MF
transition point decreases rapidly within the small interval $0\leq \kappa
\leq 1$, then saturating at a non-vanishing value in the entire strong
coupling region $\kappa >1$.

The non-vanishing value of $\left\langle \overline{\left| \Delta \right| ^{2}%
}\right\rangle $ at the MF transition point reflects smearing of the phase
transition to the SC state due to fluctuations. This effect characterizes
the GL theory of a 3D superconductor at high magnetic fields, since above
the vortex lattice melting point it can be mapped to the GL\ theory of a 1D
superconductor at zero magnetic field, where a genuine phase transition is
absent\cite{LandLif}. In the zero coupling limit, $\eta \rightarrow 0$ , the
effective GL free energy functional in a given SC layer, appearing within
the curly brackets in Eq.(\ref{fGL^n}), is equivalent to the GL energy
functional for a 0D superconductor near the SC transition at vanishing
magnetic field \cite{Ebner}: 
\begin{equation*}
f_{GL}^{\left( 0\right) }\left( \left| \psi \right| ^{2}\right) =k_{B}T_{c0} 
\left[ \frac{1}{\delta }\left( t-1\right) \left| \psi \right| ^{2}+\frac{%
0.106}{\delta }\left| \psi \right| ^{4}\right]
\end{equation*}

Here $T_{c0}=T_{c}(H\rightarrow 0)$, $t=T/T_{c0}$ , and $\delta =\left[
N\left( 0\right) vk_{B}T_{c0}\right] ^{-1}$ ( with $N\left( 0\right) $- the
electronic density of states per unit volume at the Fermi energy, and $v$-
the volume of the small SC particle), is the quantum size parameter of the
SC grain. The corresponding MF free energy is $\widetilde{f}_{GL}^{\left(
0\right) }\left( T,H\rightarrow 0\right) =-\left( \frac{9.4}{2\delta }%
\right) k_{B}T_{c0}\left( 1-t\right) ^{2}$.

For the isolated SC layer at high magnetic field and low temperature $T\ll
T_{c}\left( H\rightarrow 0\right) $ it was found that \cite{RMP01}, $\
\alpha =\frac{1}{4\hbar \omega _{c}}\ln \left( \frac{H_{c2}}{H}\right)
\approx \frac{1}{4\hbar \omega _{c}}\left( 1-h\right) $ , with $h\equiv
H/H_{c2}\left( T\rightarrow 0\right) $ , and $\beta \approx \frac{1.38}{%
\left( \hbar \omega _{c}\right) ^{2}E_{F}}$. \ Here $\omega _{c}=eH/m^{\ast
}c$ is the cyclotron frequency, with $m^{\ast }$ the in-plane effective
cyclotron mass, and $E_{F}$ the Fermi energy. The corresponding MF free
energy is $\widetilde{f}_{GL}^{\left( 0\right) }\left( T\rightarrow
0,H\right) \approx -\left( \frac{1}{16\beta _{a}}\right) E_{F}\left(
1-h\right) ^{2}$. \ For general $T$ , and $H$ values near the MF transition
line, $H_{c2}\left( T\right) =\frac{\phi _{0}}{2\pi \xi \left( T\right) ^{2}}%
=H_{c2}\left( 0\right) \left( 1-T/T_{c0}\right) $, the corresponding MF free
energy has the well known form $\widetilde{f}_{GL}^{\left( 0\right) }\left(
T,H\right) =\widetilde{f}_{GL}^{\left( 0\right) }\left( 0,0\right) \left(
1-t-h\right) ^{2}=\widetilde{f}_{GL}^{\left( 0\right) }\left( 0,0\right)
\left( 1-T/T_{c0}\right) ^{2}\left[ 1-H/H_{c2}\left( T\right) \right] ^{2}$,
so that by equating the coefficient $\widetilde{f}_{GL}^{\left( 0\right)
}\left( 0,0\right) $ in the two limiting regions of the phase boundary, i.e. 
$\widetilde{f}_{GL}^{\left( 0\right) }\left( 0,0\right) =\left( \frac{9.4}{%
2\delta }\right) k_{B}T_{c0}=\left( \frac{1}{16\beta _{a}}\right) E_{F}$,
one finds: $\delta =\frac{k_{B}T_{c0}}{1.3\times 10^{-2}E_{F}}$ with $\beta
_{a}\approx 1$.

Using the well known BCS expression for the zero temperature coherence
length, $\xi \left( 0\right) =0.18\hbar v_{F}/k_{B}T_{c0}$ , we find that 
\begin{equation}
\delta \approx 28\left( \frac{1}{k_{F}\xi \left( 0\right) }\right)
\label{delta}
\end{equation}
This enables us to estimate the effective spatial size of the SC grain in
the zero field limit. \ Since our original model system consists of a 2D SC
layer, the volume $v\equiv a^{2}$ is a 2D area, and the DOS function is that
of a 2D electron gas, $N\left( 0\right) =m^{\ast }/2\pi \hbar ^{2}=\frac{%
k_{F}^{2}}{4\pi }\frac{1}{E_{F}}$, so that $\delta \approx 35\left( \frac{%
\xi \left( 0\right) }{a}\right) \left( \frac{1}{k_{F}a}\right) $. \
Comparing this expression to Eq.(\ref{delta}) , we find that the radius of
the effective SC grain $\ a\sim \xi \left( 0\right) $ , which is
approximately equal to $a_{H_{c2}\left( 0\right) }$- the smallest length
scale in a 2D SC condensate in a magnetic field $H\approx H_{c2}\left(
0\right) $. For MgB$_{2}$ we have: \ $T_{c0}\approx 40K$ , \ $E_{F}\approx
12500K$, so that \ $\delta =0.\,23$ .

These results show that a single SC layer in a high magnetic field above the
vortex lattice melting point is equivalent to a 0D superconductor at zero
magnetic field \cite{Ebner}. The coupled layer model is therefore equivalent
to a 1D Josephson array of small SC grains, discussed, e.g. in \cite{Ebner},
or in \cite{Scalapino72}. For the relatively large effective quantum size
parameter $\delta =0.\,23$ , estimated above, the numerical simulations
reported in Ref.\cite{Ebner} show that the reduction of the transition width
by the interlayer coupling is not very significant.

To make this feature useful in our analysis of experimental data we have
approximated our result by means of an interpolation function, suggested by
Ito et al.\cite{Ito99} to fit their experimental dHvA data for a quasi 2D
organic superconductor, namely: 
\begin{equation}
\Phi _{0,interp}\left( x;\kappa \right) =x+\sqrt{\nu _{\kappa }^{2}+x^{2}}
\label{Phi_interp}
\end{equation}
where the fitting parameter $\nu _{\kappa }$ depends only on the interlayer
coupling. The equivalence of Eq.(\ref{Phi_interp}) to the interpolation
formula presented in Ref.\cite{Ito99} is made clear if we note from Eq.(\ref
{xkapbeta}) that $x=\frac{\alpha }{2\beta \beta _{a}\alpha _{0I}}=\frac{%
\Delta _{0}^{2}}{2\alpha _{I0}}\left( 1-\frac{H}{H_{c2}(T)}\right) $ , where 
$\Delta _{0}$ is the SC gap parameter at $T=0$, and $H=0$.

The best fitting function $\Phi _{0,interp}(x;\kappa )$ is represented in
Fig. (\ref{mgb2fig1}) by the dashed lines for the limiting cases $\kappa =0$
and $\kappa \rightarrow \infty $. At zero coupling the parameter $\nu
_{\kappa }$ can be obtained by comparing $\Phi _{0,interp}(x;0)$ with $\Phi
_{0}(x;0)=2\left(x+\exp (-x^{2})/\sqrt{\pi }\mathop{\rm erfc}\left(
-x\right) \right)$, to yield $\nu _{\kappa =0}=2/\sqrt{\pi }\approx 1.13$.
In the strong coupling limit the best fit is obtained for $\nu _{\kappa
=100}=.51$. Thus, the entire range of the actual smearing parameter $\alpha
_{I}=\nu _{\kappa }\alpha _{0I}$ is obtained with $\nu _{\kappa }=.51-1.13$.
\ 

The parameter $\alpha _{I}=\nu _{\kappa }\sqrt{\frac{k_{B}T}{2\beta \beta
_{a}}}$ , which controls the smearing of the phase transition by the
fluctuations, is related to the parameter $\alpha _{F}$ introduced in Ref. 
\cite{Fletcher04} (denoted there by $\alpha $ ) by: $\alpha _{F}=2\alpha
_{I}/\Delta _{0}^{2}$ , where $\Delta _{0}$ can be identified with $\Delta
_{E}$ of Ref. \cite{Fletcher04}. \ Using the above value of $\beta ,$
obtained from the 2D electron gas model, we find that: 
\begin{equation}
\alpha _{I}\approx 0.35\nu _{\kappa }\hbar \omega _{c}\sqrt{E_{F}k_{B}T}
\label{alphaI}
\end{equation}
which reflects the smearing effect due to the magnetic field, the
suppression of this smearing by the interlayer coupling parameter $\kappa $,
and the thermal smearing effect. 

For the parameters characterizing the MgB$_{2}$ data, reported by Fletcher
et al. \cite{Fletcher04}, i.e. with $m^{\ast }\approx 0.3m_{e}$ , $T=0.32K$
, $\Delta _{0}=200K$, $F=2930T$ , so that $E_{F}=\hbar \omega _{c}F/H=12464K$%
, \ our theoretical estimate is $\alpha _{I}\approx 479K^{2}$, or $\alpha
_{F}\approx 0.024$. \ This result is similar to, but somewhat smaller than
the best fitting value of $\alpha _{F}$ obtained in Ref.\cite{Fletcher04}.

This is quite a reasonable result since, as noted above, one should regard
the calculated width as a lower bound on the smearing of the SC phase
transition by all types of thermal fluctuations of the SC order parameter.
Furthermore, as clearly seen in Fig.(\ref{mgb2fig1}), the deviation of the
fluctuation effect predicted in our 3D model from the best fit to the
experimentally observed data is rather small in the magnetic field region
bellow MF $H_{c2}$, and becomes more significant at fields well above the MF
transition. \ This behavior is reasonably explained by the omission of many
fluctuation degrees of freedom in our vortex chain model \cite{ManivPhysB02}%
, which is expected to weaken progressively the fluctuation effect with
respect to the actually observed one as the magnetic field increases above
MF $H_{c2}$.

It should be also emphasized that the smearing parameter $\alpha _{I}$,
given in Eq.(\ref{alphaI}), is independent of the SC gap parameter $\Delta
_{0}$, an important parameter determining the damping of the dHvA
oscillations in the SC state \cite{RMP01}. In the fitting procedure employed
in Ref.\cite{Fletcher04} $\Delta _{0}$ was found to disagree with the SC gap
parameter $\Delta _{\pi }$ derived by other methods. \ Our estimate of the
SC fluctuations effect from the damping of the dHvA oscillation in the SC
state is thus unaffected by such a disagreement.

In conclusion, it was shown here that the transition to the low temperature
SC state in a pure, extremely type-II, 3D superconductor is smeared
dramatically by the magnetic field, comparable in magnitude to the
broadening of the transition observed in quasi 2D superconductors. \ The
dimensionality reduction by the magnetic field, responsible for this
smearing, is shown to take place in a broad field range above the vortex
lattice melting point, where the system of SC fluctuations is equivalent to
a 1D array of SC quantum dots. \ The theoretically predicted width of the
transition region is found to be in good agreement with experimental data of
the dHvA effect in the SC\ state of MgB$_{2}$. \ 

This research was supported by a grant from the Israel Science Foundation
founded by the Academy of Sciences and Humanities, and by the fund from the
promotion of research at the Technion.

\end{document}